# Tunable sub-20 fs pulses from a 500 kHz OPCPA with 15 W average power based on an all-ytterbium laser

Michele Puppin[1], Yunpei Deng[1,*], Oliver Prochnow[2], Jan Ahrens[2,3], Thomas Binhammer[2], Uwe Morgner[3], Marcel Krenz[1], Martin Wolf[1] and Ralph Ernstorfer[1,*]

[1]*Fritz-Haber Institut der Max Planck Gesellschaft, Faradayweg 4-6, 14195 Berlin, Germany*
[2]*VENTEON Laser Technologies GmbH, Holleritallee 17, D-30419 Garbsen, Germany*
[3]*Instute of Quantum Optics, Leibniz Universität Hannover, Welfengarten 1, D-30167 Hannover, Germany*
*\*Email deng@fhi-berlin.mpg.de, ernstorfer@fhi-berlin.mpg.de*

**Abstract**: An optical parametric chirped pulse amplifier fully based on Yb lasers at 500 kHz is described. Passive optical-synchronization is achieved between a fiber laser-pumped white-light and a 515 nm pump produced with a 200 W picosecond Yb:YAG InnoSlab amplifier. An output power up to 19.7 W with long-term stability of 0.3% is demonstrated for wavelength tunable pulses between 680 nm and 900 nm and spectral stability of 0.2%; 16.5 W can be achieved with a bandwidth supporting 5.4 fs pulses. We demonstrate compression of 30 µJ pulses to sub-20 fs duration with a prism compressor, suitable for high harmonic generation.

**Introduction.** In recent years, novel ultrafast spectroscopic techniques have been developed based on frequency down- and up-conversion by multiple octaves of femtosecond NIR/VIS laser pulses to the far infrared and the extreme ultraviolet (XUV) spectral range, respectively. The limited efficiencies of the conversion processes compel intense and energetic driver pulses, which so far has limited these new spectroscopies predominantly to Ti:sapphire laser systems with repetition rates in the few kHz range. In view of many time-resolved spectroscopic applications [1], in particular photoelectron spectroscopies, repetition rates of 100s of kHz represent a good compromise between high counting statistics and sufficient pulse separation to allow for full relaxation of photo-induced processes in most materials. The advent of Yb:YAG lasers providing high average powers up to the kW level [2–4] and few picoseconds pulse duration opens an alternative route to ultrashort and energetic pulses in the desired range of repetition rates through conceptually simple and power-scalable optical parametric chirped pulse amplification (OPCPA) schemes. Recently, it has been demonstrated that this approach is suitable to drive high-order harmonic generation (HHG) in gases up to MHz repetition rates [5]. Such high repetition rate sources of ultrashort XUV pulses are highly attractive for time- and angle-resolved photoelectron spectroscopy (tr-ARPES), as this reveals the evolution of the electronic structure of a crystalline material throughout the full Brillouin zone in the course of photo-induced processes [6,7]. In this type of experiments, single plateau harmonic in the energy range from 20 to 100 eV are employed for photoemission [1,7]. Since the experimental energy resolution is related to the harmonic bandwidth which in turn strongly depends on the duration, photon energy and bandwidth of the laser pulse driving HHG [8]: parametric amplification can provide the desirable adjustability of these laser parameters. Non-collinear phase matching schemes result in high gain bandwidths from the visible to the mid-infrared spectral regions. Chirped pulse amplification in this context is directly connected to gain bandwidth, due to the time gating action of the pump pulse in the nonlinear amplification process. By controlling the stretching of the seed pulse, the bandwidth can be tailored to match the pulse duration to the desired application. Here we present a novel single-stage OPCPA providing sub-20 fs pulse duration from the visible to the near-infrared with energies >30 µJ at 500 kHz repetition rate, which is an ideal tool for driving a high repetition rate XUV source for tr-ARPES experiments.

An OPCPA requires broad bandwidth seed light synchronized with energetic pump pulses. In high power systems employing a series of amplification stages in the pump arm, the synchronization of pump and seed is a major challenge as stable and efficient parametric amplification requires the timing jitter between both pulses not to exceed a few percent of the pump pulse duration. Two all-optical synchronization approaches have been demonstrated so far: the first is to seed both the OPCPA and the Yb:YAG amplifier chain with a Ti:sapphire broadband oscillator [9–11], which provides very stable seed light for the OPCPA but results in long path length differences between pump and seed; alternatively, white light continua generated from a fraction of the pump pulses may be used as seed [12–15]. The latter approach minimizes the path length differences and does not require a Ti:sapphire oscillator, but faces the challenge to generate stable white light from ~ps pulses. We follow an intermediate approach by using a master oscillator fiber amplifier (MOFA) for generating intense and stable white-light supercontinuum from few-hundred fs long pulses and for seeding a Yb:YAG InnoSlab amplifier with short path length whose frequency-doubled ps output pumps the OPCPA.

**OPCPA description.** The layout of the laser system is depicted in Fig. 1. A 25 MHz mode-locked Ytterbium fiber oscillator working in all normal dispersive regime [16] provides a spectral bandwidth of 10 nm full width half maximum (FWHM) at a central wavelength of 1030 nm. A chirped volume Bragg grating (CVBG, Optigrade

12.55 ps/nm) stretches the master oscillator pulses to about 130 ps. A dual-stage fiber preamplifier (VENTEON | PULSE : THREE PRE-AMP 3) including a fiber-coupled acousto-optical modulator is used to reduce the repetition rate in a range tunable from 0.3 to 1 MHz. The pulses are further amplified in an 80 cm long rod-type photonic crystal fiber (NKT, DC 285/100 PM-Yb-ROD) up to 9 W average power with a spectral bandwidth of 8 nm FWHM centered at 1032 nm.

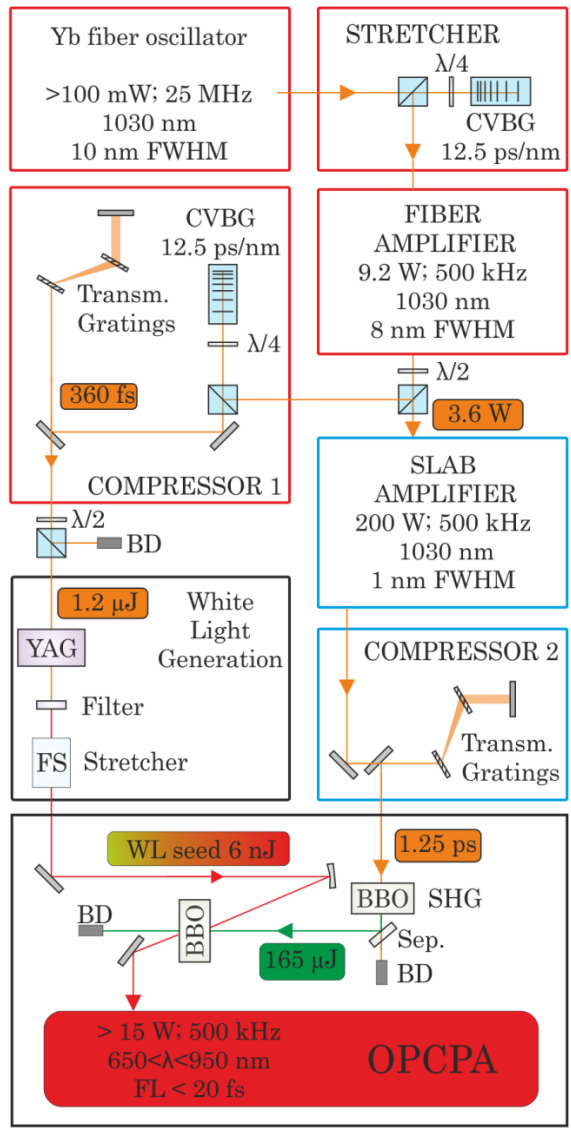

Figure 1 - OPCPA layout: CVBG = Chirped Volume Bragg Grating, FS = Fused Silica, BD = Beam Dump, Sep. = Wavelength Separator, SHG = Second Harmonic Generation.

About 40% of the MOFA output directly seeds the Yb:YAG InnoSlab amplifier (Amphos 200) with an output exceeding 200 W and a spectrum centered at 1030 nm with 1.6 nm FWHM. The combination of MOFA and InnoSlab amplifier reduces the stretching requirements for reaching hundreds of μJ compared to a pure fiber system which enables the use of an alignment-free CVBG with moderate chirp rate, instead of a more complex grating based stretcher. After compression to 1.25 ps pulse duration employing a dielectric transmission grating compressor (efficiency 75 %; gratings: LightSmyth Tech., 1000 lines/mm), the slab output is frequency doubled with 55% efficiency in a 2 mm thick BBO crystal. The resulting average power of 82.5 W at the wavelength of 515 nm is used for pumping the OPCPA.

The remaining MOFA output is compressed to 360 fs FWHM by a combination of CVBG and grating compressor and used for white light generation in YAG, which combines low filamentation threshold with high damage threshold [17]. Pulses of 1.2 μJ energy focused in a 4 mm thick YAG crystal (f = 100 mm, f# = 25) generate a smooth and stable continuum with 6 nJ pulse energy in the spectral range of 610 nm–940 nm (-10 dBc) which is used for seeding the OPCPA. We note, that a similar white light spectral power density can be obtained using a fraction of the compressed ps slab laser; to achieve the same white light power as the previous case, a higher average power (5 W) is necessary which results in long term degradation of the crystal, preventing stable WLG over hours. However, no damage of the YAG crystal is observed over many days for the sub-ps fiber laser based WLG, at the same time the continuum exhibited a better power stability.

The seed and the pump beams are overlapped non-collinearly in a 4 mm thick BBO crystal ($\theta$=24.3°, internal angle 2.4°, Type I). By stretching the seed pulses relative to the picosecond pump by transmission through approximately 60 mm fused silica glass (FS), an amplified bandwidth of 134 nm (-10 dBc) centered at 790 nm was obtained. The pump pulses are focused to a 4σ spot diameter of 975 μm; a 4σ seed mode diameter of 1100 μm was chosen to optimize the conversion efficiency. An average power output of 18.3 W was measured for the amplified signal. A walk-off compensated configuration was adopted for achieving the best beam quality; a modest parasitic second harmonic of 260 mW was measured after a dielectric dichroic mirror.

**Wavelength tunability.** One of the most attractive features of OPCPAs is the wavelength and bandwidth tunability: non-collinear phase matching allows broadband phase matching down to about 650 nm for a 515 nm pump.

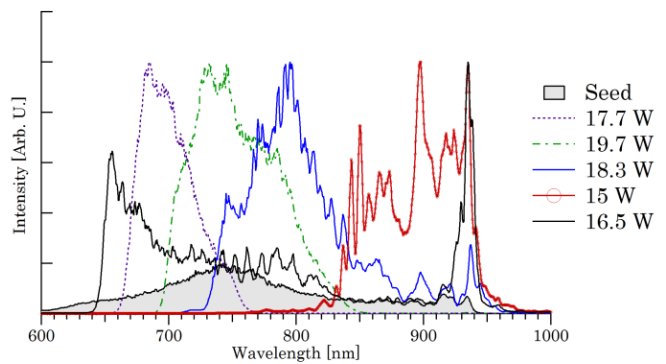

Figure 2 - OPCPA spectra illustrating wavelength and bandwidth tunability, for every spectrum the average output power is reported.

In Fig 2, the wavelength tunability of the system is reported: the output power is above 15 W when tuning the central wavelength in the 680 nm – 900 nm range. The highest output power of 19.7 W is achieved around 750 nm, close to the seed WLG maximum. Tuning the system toward the near infrared the bandwidth increases, due to lower material dispersion of the fused silica glass stretcher in this frequency domain: to achieve a comparable bandwidth around 900 nm additional 50 mm FS were added, reducing the average power to 15 W due to the lower seed level.

**Pulse characterization.** The spectrum was tuned around 790 nm and recompressed to sub-20 fs in a Brewster cut fused silica prism compressor. The pulses were characterized by SHG FROG in a 25 μm thick BBO crystal revealing a 16.8 fs FWHM temporal envelope; the results are shown in Fig. 3.

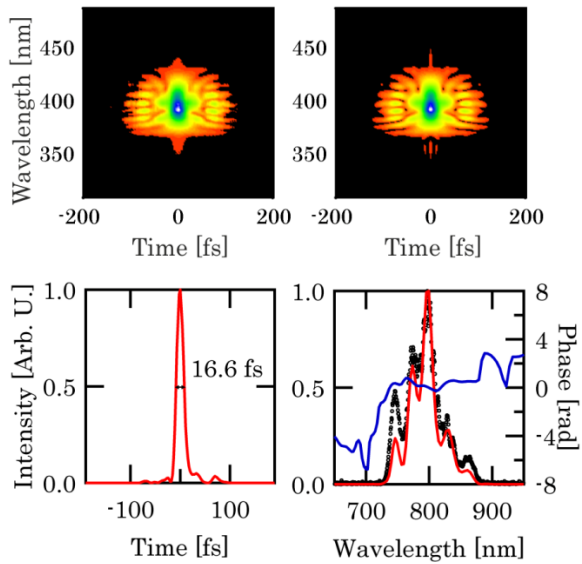

Figure 3 - Upper left panel: measured FROG trace. Upper right panel: Retrieved FROG trace. Lower left: retrieved temporal intensity showing a FWHM of 16.6 fs. Lower right: measured (black circles) and retrieved (red trace) spectral intensity.

Due to the uncompensated higher order phase, the pulse is approximately 10 % longer than the calculated Fourier limit of 15 fs. The final compressed output power is 15 W, at a repetition rate of 500 KHz with a peak power exceeding 1 GW. The same pump laser can achieve comparable output power for broad optical spectra supporting few-cycle pulses as well (see Fig. 2): by decreasing the FS transmission stretcher to 10 mm, an output with a bandwidth of approximately 300 nm (-10 dBc) at an average power of 16.5 W is demonstrated. The resulting M-shaped spectrum supports a Fourier limited pulse duration of 5.4 fs.

**Power and wavelength long-term stability.** The overall optical path length of the pump pulses in the slab amplifier and the compressor is less than 10 m and, owing to the stability of the InnoSlab design, no active synchronization techniques with the fiber-based seed are necessary. To show this, the long term average power stability was monitored over a period of 1.5 hours together with the optical spectrum (Fig. 4). For an output of 19 W, centered at 730 nm, the average power standard deviation is 0.3 %, while the spectrum centroid standard deviation is on the order of 0.2 %. This implies a timing drift between the two arms below 35 fs.

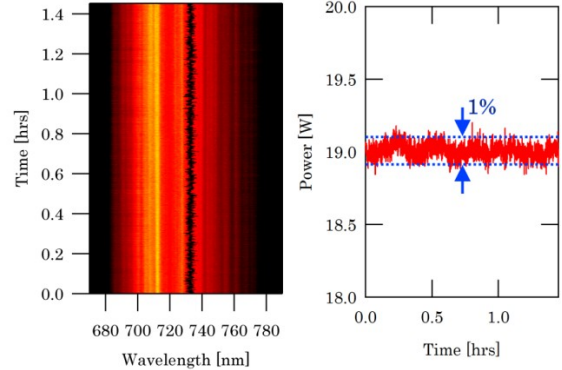

Figure 4 - Left panel: spectral intensity as a function of time; Black line: average wavelength = 732.9 ± 1.6 nm (standard deviation). Right panel: average power as a function of time = 19.00±0.05 W

**Conclusions.** In summary, a conceptually simple high power VIS/NIR OPCPA is presented. The light source is entirely based on optically synchronized Yb-lasers and the implementation of a hybrid scheme seeded by a broad white light continuum generated by a fs Yb:fiber laser system and pumped by a high power ps Yb:YAG amplifier is demonstrated. Long-term power and spectral stability are possible without active delay stabilization. Sub-20 fs, 30 μJ pulses are easily achieved with a prism compressor. We note that further scalability in term of average power is possible due to the availability of kW level Yb-pump lasers [2–4]; the implementation of a chirped mirror-based pulse compressor, as shown in [18] for a similar white light seed, promises scalability below 10 fs with high average power without active phase control. The present results are suitable for HHG at high repetition rate [19–21] and its application for XUV-based spectroscopies.

The authors acknowledge financial support by the European Union project CRONOS (grant number 280879-2).